\documentclass[conference]{IEEEtran}
\IEEEoverridecommandlockouts

\usepackage{cite}
\usepackage{multicol}

\usepackage{amsmath}
\usepackage{amsfonts}
\usepackage[T1]{fontenc}
\usepackage{algpseudocode}
\usepackage[utf8]{inputenc}
\usepackage[english]{babel}
\usepackage[usenames, dvipsnames]{color}
\usepackage{pifont}
\usepackage{booktabs}
\usepackage{mathtools}
\usepackage{textcomp}
 \usepackage{graphicx}
\usepackage{subfigure}
\usepackage[ruled,vlined,commentsnumbered]{algorithm2e}
\usepackage{amsthm}
\usepackage{setspace}

\usepackage{fancyhdr}
\fancypagestyle{firstpage}{%
  \lhead{This work has been accepted for publication in the IEEE JC\&S 2024 Conference.}
}

\SetKwInput{KwIn}{Inputs}

\theoremstyle{definition}

\theoremstyle{remark}

\theoremstyle{plain}

\newcommand{\RNum}[1]{\uppercase\expandafter{\romannumeral #1\relax}}
\newcommand{\forallK}{\forall\,k \in \mathcal{K}}

\def\BibTeX{{\rm B\kern-.05em{\sc i\kern-.025em b}\kern-.08em
    T\kern-.1667em\lower.7ex\hbox{E}\kern-.125em}}

\begin{document}

\title{Power Allocation Scheme for Device-Free Localization in 6G ISAC Networks}
\author{\IEEEauthorblockN{ Maximiliano Rivera Figueroa, Pradyumna Kumar Bishoyi, and Marina Petrova}
\IEEEauthorblockA{Mobile Communications and Computing, RWTH Aachen University, Aachen, Germany \\
Email: \{maximiliano.rivera, pradyumna.bishoyi, petrova\}@mcc.rwth-aachen.de}
}

\maketitle
\thispagestyle{firstpage}
\begin{abstract}
Integrated Sensing and Communication (ISAC) is considered one of the crucial technologies in the upcoming sixth-generation (6G) mobile communication systems that could facilitate ultra-precise positioning of passive and active targets and extremely high data rates through spectrum coexistence and hardware sharing. Such an ISAC network offers a lot of benefits, but comes with the challenge of managing the mutual interference between the sensing and communication services. In this paper, we investigate the problem of localization accuracy in a monostatic ISAC network under consideration  of inter-BS interference due to communication signal and sensing echoes, and self-interference at the respective BS. We propose a power allocation algorithm that minimizes BS's maximum range estimate error while considering minimum communication signal-to-interference-plus-noise ratio (SINR) and total power constraint. Our numerical results demonstrate the effectiveness of the proposed algorithm and indicate that it can enhance the sensing performance when the self-interference is effectively suppressed.
\end{abstract}

\section{Introduction} \label{sec:Introduction}
Going beyond pure communication services, the sixth-generation (6G) cellular networks are expected to provide localization and sensing as a service for various immersive applications \cite{Zhang2021-PMN}. Integrated sensing and communication (ISAC) emerged as a key enabling technology with an aim to integrate sensing capability into the base stations (BSs) by reusing the hardware and spectrum resources \cite{liu2022}. However, a single ISAC-enabled BS is unlikely to meet ubiquitous and accurate sensing and positioning requirements due to its limited resources (e.g., bandwidth, power, sensing time, field of view). Hence, ITU-R has proposed the notion of an \textit{ISAC network} in its recent report \cite{ITU2022}, which entails the cooperation of multiple ISAC-enabled BSs and utilization of the existing dense cellular infrastructure to provide precise sensing services and improved target localization across a large contiguous area.


Despite the benefits of the ISAC network, eliminating the mutual interference between ISAC BSs is a challenging issue. There are two types of interference in the ISAC network. First, at the network level, there is inherent inter-BS interference caused by communication and echo signals \cite{xin2023mag}. Second, at each BS, there occurs the issue of self-interference. When the BS performs monostatic sensing, it receives a delayed version of the transmitted signal as an interference to the receiving echo signal. Developing resource allocation schemes specifically tailored for ISAC networks while considering both types of interference and characterizing the impact of these interferences on both sensing and communication performance is still an open issue. Further, in the context of device-free (passive) object localization, the issue is even more challenging since the interference increases the range estimation error of each BS while degrading the overall localization accuracy. To this end, we seek to address the following questions pertaining to ISAC networks: i) \textit{To what extent do self-interference and network-level interference impact the accuracy of device-free localization?}, and ii) \textit{how to improve the overall localization accuracy taking communication constraints into account?}
%

In recent literature, most of the works focus on improving the performance of single-cell ISAC systems \cite{liu2022,Chiriyath2016} and very handful of works focused on the ISAC network \cite{Huang2022,jiang_VTC_2022,zinat_kth_2022}. The authors in \cite{Huang2022} proposed a coordinated power allocation strategy for bistatic sensing in an ISAC network. The proposed scheme minimizes the overall system power while considering the minimum signal-to-interference-noise (SINR) constraint at communication user and maximum Cram$\acute{\text{e}}$r-Rao lower bound (CRLB) for the single target localization. Although CRLB captures the theoretical lower bound on the variance of the unbiased range estimator, it ignores the impact of the range estimation error of individual BS on the overall localization error. The authors in \cite{jiang_VTC_2022} analyzed the effect of mutual interference between two ISAC BSs and proposed a collaborative precoding design for interference elimination while adhering to both total power and per antenna power constraints. However, the communication users are modelled as sensing targets in the considered scenario. As a result, the suggested technique may not be feasible in situations where the communication user and sensing target are distinct entities situated in separate locations. In contrast to traditional cellular architecture for ISAC, authors in \cite{zinat_kth_2022} explored the utilization of a cell-free massive MIMO system for the purpose of multistatic sensing of a single target. A power allocation scheme is proposed to maximize the probability of detection of the target while considering communication user SINR and total power constraints into account. The proposed scheme, however, does not consider any effect of self-interference and thus is not applicable to a monostatic sensing-based ISAC network.

The goal of this paper is to investigate and propose a power allocation method that achieves the highest possible device-free localization accuracy in monostatic sensing-based ISAC networks, while meeting the performance of the communication service. Our analysis takes into account inter-BS interference and self-interference. The communication performance is characterized by the downlink SINR attained by the communication user. For characterizing the localization performance, we aim to minimize the maximum range estimate error of a set of BSs. The range estimation of all participating BSs is crucial in determining the true location of the object. A high range estimation error in one of the BSs will result in a degradation of the overall localization accuracy. Therefore, our aim is to select the BS that exhibits the greatest range estimation error and enhance its sensing SINR in order to reduce that error. This also ensures fairness inside the system. Thereafter, we formulate a power allocation problem with the objective of improving localization performance while considering minimum communication SINR and total power constraint. The main \textit{contributions} of the work are as follows:

\begin{itemize}
    \item We formulate a min-max optimization problem aimed at enhancing the accuracy of range estimates for the least-performing BS in the network. Since the formulated problem is nonlinear and non-convex, first, we added relaxations to transform it into a convex problem and thereafter, proposed an iterative algorithm to obtain the solution.
    \item We evaluate the performance of the proposed power allocation scheme in terms of localization accuracy through extensive simulations. We investigate the influence of self-interference on the standard deviation (STD) of range measurements. Additionally, we examine the impact of power allocation for high levels of self-interference.
\end{itemize}

The rest of the paper is organized as follows. We describe the system model in Section \ref{sec:SystemModel}. In Section \ref{sec:PowerAllocation}, we describe the formulated optimization problem and explain the proposed algorithm. Further, we show the performance evaluation in Section \ref{sec:PerformanceEvaluation} and conclude the paper in Section \ref{sec:Conclusions}.


\section{System Model}\label{sec:SystemModel}
\begin{figure}[t]
    \centering
    \includegraphics[width=0.8\linewidth, trim={0cm 0cm 0cm 0cm},clip]{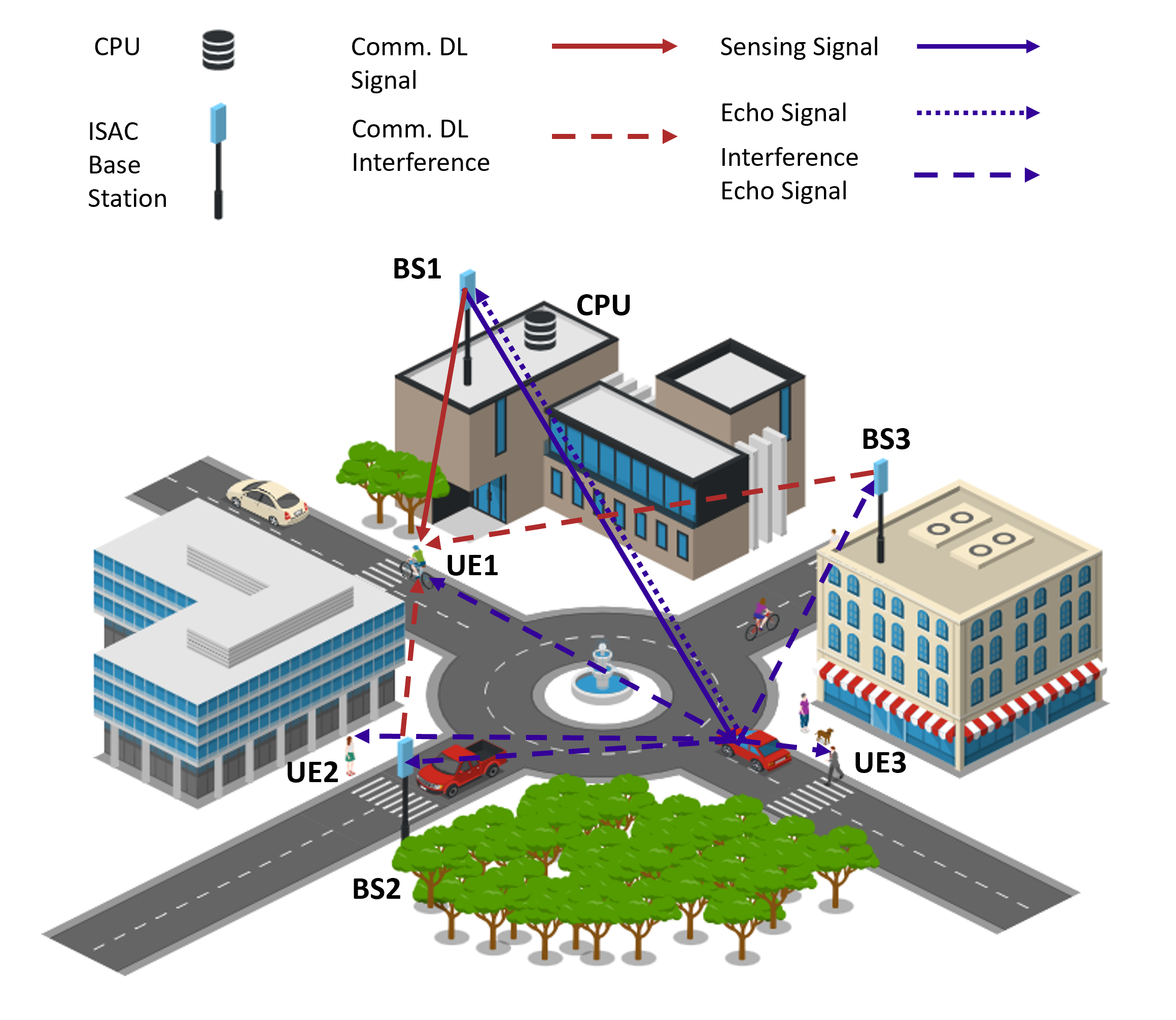}
    \caption{Illustration of an ISAC network where the BSs perform both communication and sensing in an urban scenario. The figure particularly depicts the DL communication from BS1 to UE1, where the latter suffers interference from other BSs, and the sensing from BS1 causes interference to other BSs and UEs.}
    \label{fig:illustration_SystemModel}
    \vspace{-0.45cm}
\end{figure}

We consider an ISAC network consisting of $K$ BSs in an urban area, where each BS possesses communication and sensing capabilities, as shown in Fig.~\ref{fig:illustration_SystemModel}. For sensing services, each BS acts as a monostatic radar with full-duplex capabilities \cite{figueroa2023cooperative}. Further, all the BSs are synchronized and connected to a central processing unit (CPU) via a high-capacity backhaul link. In this system, each BS operates in frequency range 1 (FR1) and serves a unique user equipment (UE) by sending individual messages. The $K$ BSs aim to, while performing downlink (DL) communication with their UEs, locate a passive target of unknown 3D position $\boldsymbol{x} = [x,\,y,\,z]^T$ moving with an unknown speed of $\boldsymbol{v}$ m/s. The location of the BSs and UEs are known to the system and are denoted by $\boldsymbol{x}_{\text{BS},\,k} = [x_{\text{BS},\,k},\,y_{\text{BS},\,k},\,z_{\text{BS},\,k}]^T$ and $\boldsymbol{x}_{\text{UE},\,\check m} = [x_{\text{UE},\,\check m},\,y_{\text{UE},\,\check m},\,z_{\text{UE},\,\check m}]^T$, with $k,\,\check m \in \mathcal{K}$, and $\mathcal{K} = [1,\dots,K]$. Under the conditions above, the $k$-th BS sends an individual message to the $\check k$-th UE. For clarity, the indices that contains a check, e.g., $\check m$, are related to UEs, thus, $\boldsymbol{x}_{\text{BS},\, k} = \boldsymbol{x}_{ k}$ and $\boldsymbol{x}_{\text{UE},\,\check m} = \boldsymbol{x}_{\check m}$.

In order to analyze the sensing and communication performance, we initially examine the stand-alone systems before proceeding to evaluate the integrated system.

\subsection{Communication-only Analysis}
We begin our analysis by examining Fig.~\ref{fig:illustration_SystemModel} in the absence of the target. Each BS performs DL communication to their UE, where the received signal at each UE suffers from interference from other BSs' DL signals.
Let $s_k(t)$ denote the communication DL transmit signal based on an orthogonal frequency division multiplexing (OFDM) waveform from the $k$-th BS. The DL signal consists of $N_s$ active subcarriers with $\Delta f$ of subcarrier spacing (SCS) and $M_s$ OFDM symbols. The baseband time-domain signal is represented as follows:
\begin{equation}
    s_k(t) = \sum_{m^{\prime}=0}^{M_s-1} \sum_{n^{\prime}=0}^{N_s-1} c_{m^{\prime},\,n^{\prime}}^k\, e^{j\,2\pi n^{\prime} \Delta f\,t} g(t - m^{\prime}T),\, \forallK \label{eq:ofdmSignal}
\end{equation}
where $c_{m^{\prime},\,n^{\prime}}^k$ represent the complex modulation OFDM symbols for a given symbol $m^{\prime}$ and subcarrier $n^{\prime}$ of the $k$-th BS, and $g(\cdot)$ is the pulse shaping function. The overall OFDM symbol duration consists of $T = T_{cp} + T_s$, where $T_{cp}$ is the cycle prefix duration and $T_s$ is the symbol duration. We assume that the signal $s_k(t)$ is ergodic and independent from each other. Further, we assume that $\mathbb{E} \{s_k(t)\} = 0,\, \forallK$ and has unit variance \cite{Huang2022}.
Let $\rho_k$ denote the transmit power of the $k$-th BS with $0 \leq \rho_k \leq P_{\text{max}}, \forallK$, where $P_{\text{max}}$ is the maximum power that any BS can use. Then, the transmitted signal from the $k$-th BS is
\begin{equation}
    u_k(t) = \sqrt{\rho_k}\,s_k(t),\, \forallK \label{eq:TxBBSignal}
\end{equation} 
Let $h_{\check{m}, k}$ denote the channel coefficient from the $k$-th BS to the $\check m$-th UE. It is modelled as a Rician channel whose line-of-sight (LoS) component is deterministic and non-LoS (NLoS) component follows a zero-mean and unit-variance circular symmetric complex Gaussian (CSCG) random variable \cite{jiang_VTC_2022}. Then, the received signal at the $\check m$-th UE is
\begin{align}
&y_{\check m}(t) = \sum_{k=1}^{K} h_{\check{m}, k}\,\sqrt{\rho_k}\,s_k(t) + n_{\check m}(t)\nonumber\\
& = \underbrace{h_{\check{m}, m}\,\sqrt{\rho_m}\,s_m(t)}_{\text{Desired signal}} + \underbrace{\sum_{k=1,\, k\neq m}^{K} h_{\check{m}, k}\,\sqrt{\rho_k}\,s_k(t)}_{\text{Interference signal due to other BSs' DL}} + \underbrace{n_{\check m}(t)}_{\text{Noise}} \label{eq:CommRxSignal}
\end{align}
where the measured noise at the $\check m$-th UE is represented by $n_{\check m}(t) \sim \mathcal{CN}(0,\, \sigma_{\check m}^2)$, with $\sigma_{\check m}$ the standard deviation (STD). The SINR at the $\check m$-th UE is given by
\begin{equation}
    \gamma_{\check m}^{\text{comm}}(\boldsymbol{\rho}) = \frac{\rho_{ m} \,|h_{\check{m}, m}|^2}{\sum_{k=1,\, k\neq m}^{K} \rho_k |h_{\check{m}, k}|^2\,+\,\sigma_{\check m}^2} 
\end{equation}
where $\boldsymbol{\rho} = [\rho_1,\dots,\rho_K]^T$ is the power vector. 

\subsection{Sensing-only Analysis}
We consider the same scenario presented in Fig.~\ref{fig:illustration_SystemModel} but without UEs. The BSs aim to estimate their relative distances to a passive target of unknown location and speed. Each BS transmits the signal given by Eq. \eqref{eq:TxBBSignal} and captures its echo reflected by the target. Further, the received echo is interfered with by the bistatic echoes from the other BSs, and its own self-interference (SI) \cite{Temiz2022}. Therefore, the received signal at the $n$-th BS, $n \in \mathcal{K}$, is 
\begin{align} r_n(t) &= \sum_{k = 1}^K \gamma_{n,\,k}\,\sqrt{\rho_k}\,s_k(t - \tau_{n, k})  +\beta_n \sqrt{\rho_n}\,s_n(t-\tau_{n, \ell})\nonumber\\ &
+w_n(t) \nonumber\\
=&\, \underbrace{\gamma_{n, n}\,\sqrt{\rho_n} \,s_n(t - \tau_{n, n})}_{\text{Desired echo signal}} + \underbrace{\sum_{k = 1,\,k\neq n}^K \gamma_{n,\,k}\,\sqrt{\rho_k}\,s_k(t - \tau_{n, k})}_{\text{Interference signal due to other BSs echoes}} \nonumber\\
&+\underbrace{\beta_n\sqrt{\rho_n}\,\,s_n(t-\tau_{n, \ell})}_{\text{Self-interference}} + \underbrace{w_n(t)}_{\text{Noise}}  \label{eq:SensingRxSignal}
\end{align}
where $\gamma_{n, k}$ is a coefficient that gathers the effect of the target radar cross section (RCS) and the signal propagation path-loss from the $k$-th BS to the $n$-th BS through a reflection from the target \cite{Richard2004}. The delay $\tau_{n, k}$ represents the time travelled by the signal, and it is given by $\tau_{n, k} = (||\boldsymbol{x}_k - \boldsymbol{x}||_2 + ||\boldsymbol{x} - \boldsymbol{x}_n||_2)/c$,
and $\beta_n$ and $\tau_{n, \ell}$ represent the attenuation and delay time due to the SI in the $n$-th BS. Further, $w_n(t)$ is the measured noise at the $n$-th BS, represented by $w_n(t) \sim \mathcal{CN}(0,\, \sigma_{n}^2)$. Recall that $\sigma_n \neq \sigma_{\check n},\,\forall n,\,\check n \in \mathcal{K}$, where the former is the noise STD in the receiver of the $n$-th BS and the latter is the noise STD in the receiver of the $\check n$-th UE. It is important to ensure that the SI does not exceed the saturation limit of the Analog-to-digital converters (ADCs) and receiver amplifiers, which can prevent full duplex operation and digital SI cancellation \cite{Temiz2022}. Then, the sensing SINR is given by
\begin{equation}
    \gamma_{n}^{\text{sensing}}(\boldsymbol{\rho}) = \frac{\rho_{ n} \,|\gamma_{n, n}|^2}{\sum_{k=1,\, k\neq n}^{K} \rho_k |\gamma_{n, k}|^2\,+\rho_n\,|\beta_n|^2+\,\sigma_{ n}^2} 
\end{equation}

SINR expressions $\gamma_{\check m}^{\text{comm}}(\boldsymbol{\rho})$ and $\gamma_{n}^{\text{sensing}}(\boldsymbol{\rho})$ assume no target presence and no communication link to a UE, respectively. However, communication and sensing coincide in the presented ISAC network, in which extra interference is added to the received signal.
\subsection{Integrated communication and sensing analysis}

The received signals are modified by an extra term due to the other service, causing ISAC interference. In the case of the received DL signal in the $\check m$-th UE, Eq. \eqref{eq:CommRxSignal} is modified  as follows
\begin{equation}
    y_{\check m}^{ov}(t) = y_{\check m}(t) + \sum_{k=1}^K \gamma_{\check m,\, k}\,\sqrt{\rho_k} \, s_k(t - \tau_{\check m,\, k}) \label{eq:CommRxSignalISAC}
\end{equation}
where $\gamma_{\check m, \,k}$ and $\tau_{\check m,\, k}$ are equivalent to the one defined in Eq. \eqref{eq:SensingRxSignal} but the signal arrives at the $\check m$-th UE. In particular, the delay is $    \tau_{\check m, k} = (||\boldsymbol{x}_k - \boldsymbol{x}||_2 + ||\boldsymbol{x} - \boldsymbol{x}_{\check m}||_2)/c$.
Recall that $\boldsymbol{x}_{\check m} = \boldsymbol{x}_{\text{UE},\,\check m}$. Further, the received signal in Eq. \eqref{eq:CommRxSignalISAC} can be regrouped to
\begin{align} &y_{\check m}^{ov}(t) = \underbrace{h_{\check{m}, m}\,\sqrt{\rho_m}\,s_m(t) + \gamma_{\check m,\, m}\,\sqrt{\rho_m} \, s_m(t - \tau_{\check m,\, m})}_{\text{Desired signal} \footnotemark }\nonumber\\ 
&+ \underbrace{\sum_{k=1,\, k\neq m}^{K} h_{\check{m}, k}\,\sqrt{\rho_k}\,s_k(t) + \gamma_{\check m,\, k}\,\sqrt{\rho_k} \, s_k(t - \tau_{\check m,\, k})}_{\text{Interference signal due to other BSs DL and bistatic echoes}} + \underbrace{n_{\check m}(t)}_{\text{Noise}} 
\end{align} 
where $y_{\check m}^{ov}$ stands for the overall received signal in the $\check m$-th UE.
\footnotetext{The second term of the desired signal is included as desired signal since it is another propagation path to the UE, which is not accounted for in the communication model.}
Then, the overall communication SINR is given by 
\begin{equation}
    \gamma_{\check m}^{\text{comm},\,ov}(\boldsymbol{\rho}) = \frac{\rho_{ m} \,|h_{\check{m}, m}|^2 + \rho_{ m} \,|\gamma_{\check{m}, m}|^2}{\sum_{k=1,\, k\neq m}^{K} \rho_k (|h_{\check{m}, k}|^2+|\gamma_{\check{m}, k}|^2)\,+\,\sigma_{\check m}^2} \label{eq:SINR_Comm_Ov}
\end{equation}
Analogously, the received echo is distorted by the communication DL signals. Following Eq. \eqref{eq:SensingRxSignal}, the overall received signal in the $n$-th BS is given by
\begin{align} 
&r_n^{ov}(t) = \,r_n(t) + \sum_{k=1, \,k\neq n}^K h_{n,\,k} \sqrt{\rho_k}\, s_k(t)
\nonumber\\
 &=\,\underbrace{\gamma_{n, n}\,\sqrt{\rho_n} \,s_n(t - \tau_{n, n})}_{\text{Desired signal}} + \underbrace{\beta_n\sqrt{\rho_n}\,\,s_n(t-\tau_{n, \ell})}_{\text{Self-interference}} +\nonumber\\
& \underbrace{\sum_{k = 1,\,k\neq n}^K \sqrt{\rho_k}(\gamma_{n,\,k}\,s_k(t - \tau_{n, k})+ h_{n,\,k}\, s_k(t))}_{\text{Interference signal due to other BSs bistatic echoes and DL signals}} + \underbrace{w_n(t)}_{\text{Noise}} \label{eq:SensingRxSignalISAC}
\end{align}
where $h_{n, k}$ is the channel between two different BSs.
Then, the overall sensing SINR is
\begin{align}
    &\gamma_{n}^{\text{sensing},\,ov}(\boldsymbol{\rho}) = \nonumber\\
    &\frac{\rho_{ n} \,|\gamma_{n, n}|^2}{\sum_{k=1,\, k\neq n}^{K} \rho_k (|\gamma_{n, k}|^2 + |h_{n,\,k}|^2)\,+\rho_n\,|\beta_n|^2+\,\sigma_{ n}^2} \label{eq:SINR_Sens_Ov}
\end{align}


 From  Eqs. \eqref{eq:SINR_Comm_Ov} and \eqref{eq:SINR_Sens_Ov}, it can be seen that adding sensing to communication or communication to sensing leads to a mixed term in the SINR expression, where the interference component is higher than the stand-alone version. The effect of interference on overall system performance is discussed in the subsequent sections.

\section{Power Allocation Scheme in ISAC networks}\label{sec:PowerAllocation}

In this section, first, we formulate the power allocation problem for the considered ISAC network. Thereafter, discuss the solution approach in detail and propose an iterative algorithm to compute the optimal power level. 

\subsection{Problem Formulation} \label{sec:problemformulation}
In order to formulate the power allocation problem, we consider some relevant performance metrics for communication and sensing.
On the one hand, to ensure a communication services, the SINR expressed in Eq. \eqref{eq:CommRxSignalISAC} needs to be greater than a particular threshold. Otherwise, the system cannot detect and decode the signal from the noise level. 
On the other hand, the sensing performance can be assessed by the STD of the range measurements, which is given by \cite{Richard2004}
\begin{equation}
    \sigma_{R,\,n}(\boldsymbol{\rho}) = \frac{c}{2\,BW_n\,\sqrt{2\,\gamma_{n}^{\text{sensing},\,ov}(\boldsymbol{\rho}) }} \label{eq:stdRange}
\end{equation}
where $c$ is the speed of light and $BW_n$ the bandwidth used in the $n$-th BS. Therefore, the sensing performance of each BS will be different depending on the bandwidth used and the measured SINR. 

Furthermore, we are also interested in improving the performance of the sensing services of the network while ensuring a minimum communication SINR. Specifically, we aim to improve the worst-sensing performer in the network in order to ensure good-quality measurements from all participating BSs carrying out sensing.
Consequently, we formulate the optimization problem as:
\begin{align}
    \min_{\boldsymbol{\rho}} \quad & \max_{k \in\mathcal{K}} \quad \sigma_{R,\,k}(\boldsymbol{\rho}) & \tag{P1} \label{P1} \\
    \textrm{s.t.} \quad & \gamma_{\check m}^{\text{comm},\,ov}(\boldsymbol{\rho}) \geq \Gamma_{\check m}^{\text{comm}}, \; &\forall \,\check m \in \mathcal{K} \tag{P1.a} \label{P1a}\\
	\quad & 0 \leq\rho_n \leq P_{\text{max}}, &\forall\, n \in \mathcal{K} \tag{P1.b} \label{P1b}
\end{align}
where the minimization is over the worst STD range estimation of all BSs. \eqref{P1a} stands for the constraint that the communication SINR has to be above a certain level, and \eqref{P1b} ensures feasible power constraints.
The objective function of problem \eqref{P1} is nonlinear and non-convex. It is nonlinear due to the square root in Eq. \eqref{eq:stdRange}, and the non-convexity is due to the coupling of variables ($\boldsymbol{\rho}$) in the SINR expressions. 
Therefore, \eqref{P1} is a nonlinear non-convex semidefinite program \cite{boydconvex}.

\subsection{Proposed Solution}\label{sec:proposedmethod}

In order to solve problem \eqref{P1}, we first find an equivalent optimization problem. We start by replacing the STD of the range measurements with the respective SINR, where minimizing the former is equivalent to maximizing the latter. Moreover, as the square root function is monotone and assuming equal bandwidth allocation, the equivalent problem is as follows:
\begin{align}
    \max_{\boldsymbol{\rho}} \quad & \min_{k \in \mathcal{K}} \quad \gamma_{k}^{\text{sensing},\,ov}(\boldsymbol{\rho}) & \tag{P2} \label{P2}\\
    \textrm{s.t.} \quad & \eqref{P1a},\eqref{P1b} \nonumber
\end{align}
where the accuracy of the range estimation is replaced by the sensing SINR. Further, the minimization problem becomes a maximization problem.
Problem \eqref{P2} is still non-convex due to the objective function.
To solve \eqref{P2}, we add a slack variable $\eta \geq 0$ and relax the optimization problem \cite{boydconvex},
\begin{align}
    \max_{\boldsymbol{\rho},\,\eta} \quad &  \eta & \tag{P3} \label{P3}\\
    \textrm{s.t.} \quad & \eta \leq \gamma_{n}^{\text{sensing},\,ov}(\boldsymbol{\rho}), &\forall\, n \in \mathcal{K} \tag{P3.a} \label{P3a}\\
 \quad & \eqref{P1a},\eqref{P1b} \nonumber
\end{align}
The objective function of \eqref{P3} is convex. However, constraint \eqref{P3a} is non-convex due to the coupling in the variables $\eta$ and $\boldsymbol{\rho}$. To tackle this problem, we follow a similar approach as in \cite{Yang2022}. Let $\xi_k\geq 0,\forall k \in \mathcal{K}$ a slack variable, then, constraint (P3.c) can be written as
\begin{equation}
    \xi_n \geq \sum_{k=1,\, k\neq n}^{K} \rho_k (|\gamma_{n, k}|^2 + |h_{n,\,k}|^2)\,+\rho_n\,|\beta_n|^2+\,\sigma_{ n}^2,\,\forall\, n \in \mathcal{K} \label{eq:xi_denominator}
\end{equation}
and
\begin{align}
    \rho_{ n} \,|\gamma_{n, n}|^2 &\geq \eta \cdot\xi_n \nonumber\\
    &= \frac{1}{4}\Big((\xi_n+\eta)^2 - (\xi_n-\eta)^2\Big),\,\forall\, n \in \mathcal{K} \label{eq:eta_xi_numerator} 
\end{align}
From these relaxations, Eq. \eqref{eq:xi_denominator} is convex. However, Eq. \eqref{eq:eta_xi_numerator} is still non-convex. Therefore, we apply the difference of convex (DC) functions \cite{boydconvex} on the right side to convexify it. The approximation of Eq. \eqref{eq:eta_xi_numerator} is given by
\begin{align}
    \rho_{ n} \,|\gamma_{n, n}|^2 &\geq \frac{1}{4}\Big[(\xi_n+\eta)^2 - 2(\xi_n^{(i-1)} - \eta^{(i-1)}) (\xi_n - \eta) \nonumber\\
    &+ (\xi_n^{(i-1)} - \eta^{(i-1)}) ^2\Big],\,\forall\, n \in \mathcal{K}\label{eq:eta_xi_iterative_numerator}
\end{align}
where the superscript indicates the value of the parameter one iteration before. Thus, the optimization problem becomes
\begin{align}
    \max_{\boldsymbol{\rho},\,\eta,\,\boldsymbol{\xi}} \quad &  \eta & \tag{P4} \label{P4}\\
    \textrm{s.t.} \quad & \eqref{P1a},\eqref{P1b},\eqref{eq:xi_denominator},\eqref{eq:eta_xi_iterative_numerator}\nonumber
\end{align}
where $\boldsymbol{\xi} = [\xi_1, \dots, \xi_K]^T$. Problem \eqref{P4} is an iterative convex optimization problem that can be solved by using optimization solvers such as CVX \cite{cvx}. The Algorithm \ref{algo:PowerAllocationOpt} depicts the complete algorithm steps, where $\boldsymbol{\xi}^{(0)}$ and $\eta^{(0)}$ are the initial conditions for the slack variables, $\boldsymbol{\Gamma}^{\text{comm}} = [\Gamma_{\check{1}}^{\text{comm}},\dots, \Gamma_{\check{K}}^{\text{comm}}]^T$ is the communication threshold vector, and $\epsilon$ is the convergence threshold.

\textit{Complexity Analysis}: The complexity of solving \eqref{P1} is given by the number of iterations needed by the successive convex approximations (SCA) to converge and the complexity of \eqref{P4} \cite{boydconvex,cvx}. The former is given by $\mathcal{O}(\sqrt{4K} \log_2(1/\epsilon))$, where the term $4K$ is due to the number of constraints, and the latter is given by $\mathcal{O}((2K+1)^2\,\cdot 4K)$, where the term $2K+1$ is the number of variables. Therefore, the complexity required to solve problem \eqref{P1} is $\mathcal{O}(K^{3.5}\log_2(1/\epsilon))$. 

\vspace{-0.2cm}
\begin{algorithm}
	\small
	\SetKwInOut{Input}{Inputs}
	\SetKwInOut{Output}{Outputs}
	\Input{$\boldsymbol{\xi}^{(0)}$, $\eta^{(0)}$, $\boldsymbol{\Gamma}^{\text{comm}}$, $P_{\text{max}}$,  $\epsilon$}
	\Output{ $ \boldsymbol{\rho}^{*}$, $\boldsymbol{\xi}^*$, $\eta^*$}
	Set $ converge=0, i=0$\\
	\While{converge = 0}{
		Solve \eqref{P4} for $\big(\boldsymbol{\xi}^{(i)},\,\eta^{(i)},\,\boldsymbol{\Gamma}^{\text{comm}},\, P_{\text{max}} \big)$,\\
		and get the solution for $\big(\boldsymbol{\xi}^{(i+1)},\,\eta^{(i+1)},\, \boldsymbol{\rho}^{(i+1)}\big)$\\
        $i \leftarrow i+1$\\
		\uIf{$|\eta^{(i)} - \eta^{(i-1)}|<\epsilon$}{$converge \leftarrow 1$}
	}
	\caption{Power Allocation Algorithm}
	\label{algo:PowerAllocationOpt}

\end{algorithm}

\vspace{-0.2cm}
\section{Performance Evaluation}\label{sec:PerformanceEvaluation}
This section outlines the setup and scenario used to validate our system model and assess our proposed solution. Section \ref{sec:SetupScenario} details the implemented scenario and parameters used, while Section \ref{sec:SimResults} presents and analyzes the results.

\begin{figure}[t]
    \centering
    \includegraphics[width=0.8\linewidth, trim={0cm 0cm 0cm 0cm},clip]{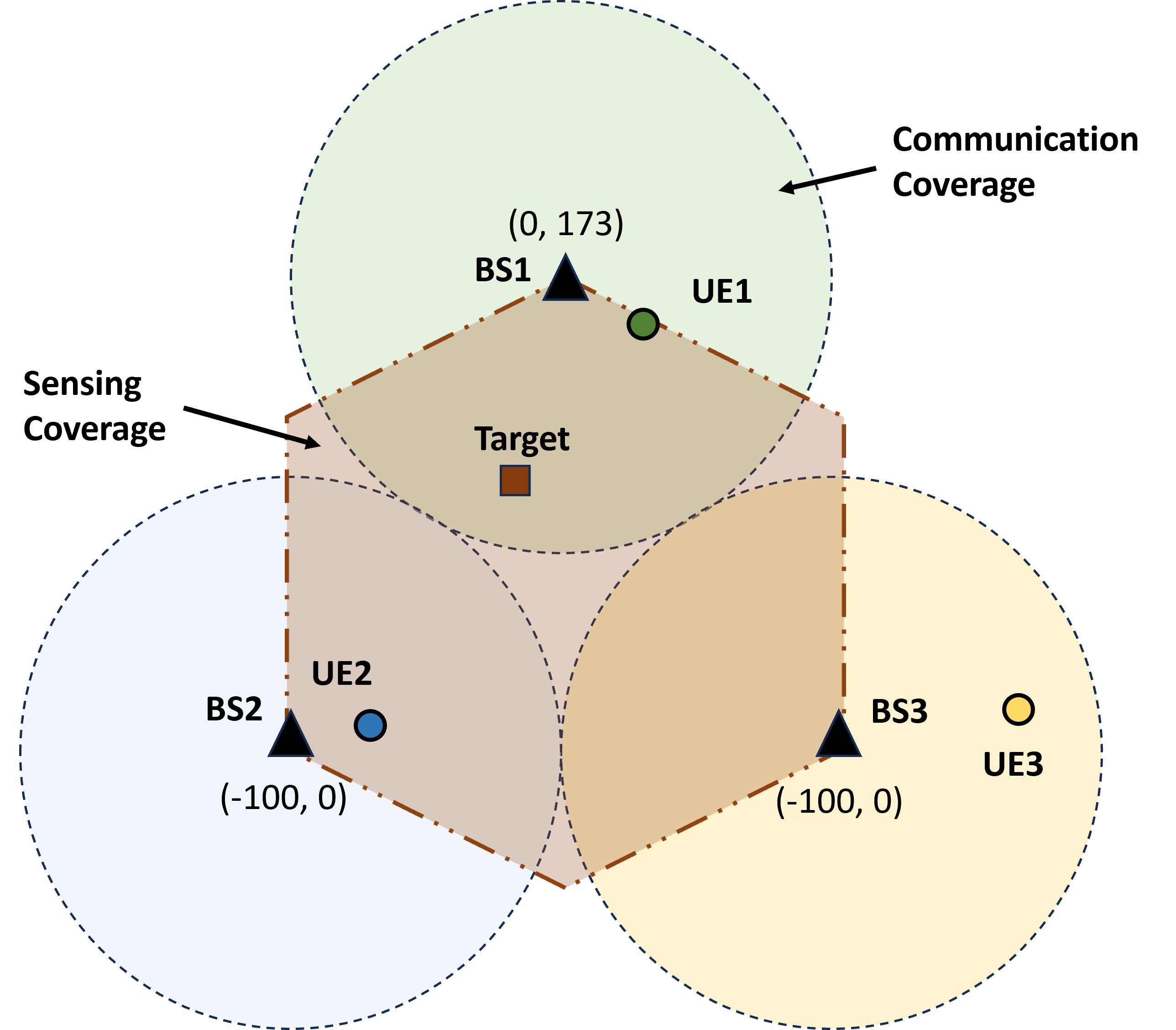}
    \caption{Snapshot of one realization of the studied scenario in the power allocation optimization problem. This figure depicts three BSs and their respective UEs in a color-code pattern. Each BS possesses a communication coverage displayed by a dashed circle, whereas the ISAC network comprises a sensing coverage indicated by the brown hexagon. All values are in meters.}
    \label{fig:Scenario}
    \vspace{-0.40cm}
\end{figure}

\subsection{Setup and Scenarios}\label{sec:SetupScenario}

To evaluate the performance of the proposed algorithm, we use the ISAC network presented in Fig. \ref{fig:Scenario}. This figure depicts the deployment of three BSs, three UEs, and one target. Each BS possesses its communication coverage, illustrated as a dashed circle, and communicates with its respective UE represented by the color code. For instance, BS1 sends DL communication to UE1. The hexagon indicates the sensing coverage of the network.
In order to assess the proposed algorithm, we implement a Monte-Carlo simulation, where at each different realization, the UEs are uniformly randomly placed within their communication coverage, and the target is uniformly randomly located within the sensing coverage. Therefore, Fig. \ref{fig:Scenario} depicts a snapshot of the simulated scenario. 
In this scenario, each BS is placed at an altitude of $10$ m, whereas the UEs and target have an altitude of $1$ m. All BSs are deployed with sectorized antennas of $120^{\circ}$, forming the sensing coverage, and the target is modelled with $7$ dBsm. The OFDM signal is composed of $M_s =28$ OFDM symbols and $N_s = 3264$ active subcarriers with a carrier frequency of $3.5$ GHz. The subcarrier spacing is set to $30$ kHz and $BW_i = 100$ MHz, $\forall\,i \in \mathcal{K}$. The maximum power available at each BS is $23$ dBm, the self-interference (SI) level and communication threshold are set to $-90$ dBm and $0$ dB, respectively, unless stated otherwise. The communication channel is modelled as a Rician with $K$-factor of $5$ dB \cite{jiang_VTC_2022} and a path-loss exponent of 2.5, whereas the sensing channel is modelled as a Line-of-sight model with a path-loss exponent of 2. 

\subsection{Simulation Results and Discussion}\label{sec:SimResults}
\begin{figure}[t]
    \centering
    \includegraphics[width=0.8\linewidth, trim={2cm 6.5cm 2cm 7cm},clip]{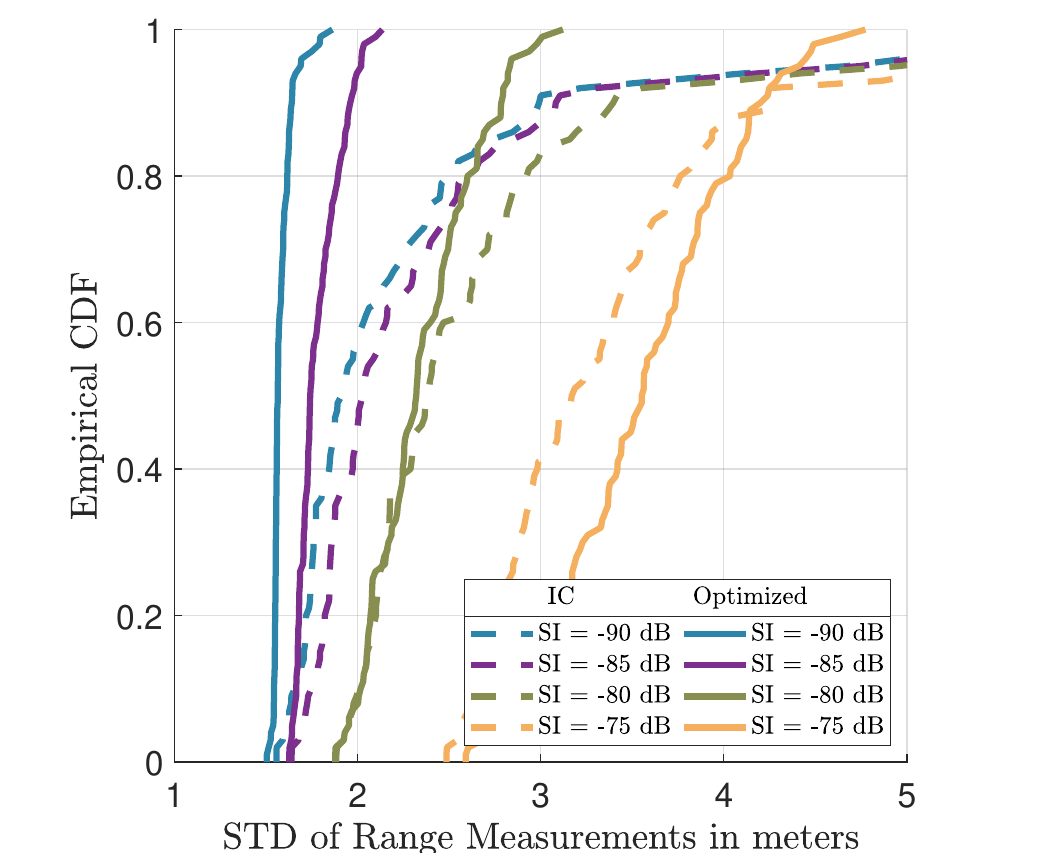}
    \caption{Empirical cumulative density function vs. the STD of range measurements for different self-interference levels. The initial conditions (IC) are computed using $23$ dBm for all BSs, and $\Gamma^{\text{comm}}_{\check m} = 0,\,\forall \check m \in \mathcal{K}$.}
    \label{fig:CDF_v_STD_underSI}
    \vspace{-0.4cm}
\end{figure}
\begin{figure}[t]
    \centering
    \includegraphics[width=0.8\linewidth, trim={2cm 6.5cm 2cm 7cm},clip]{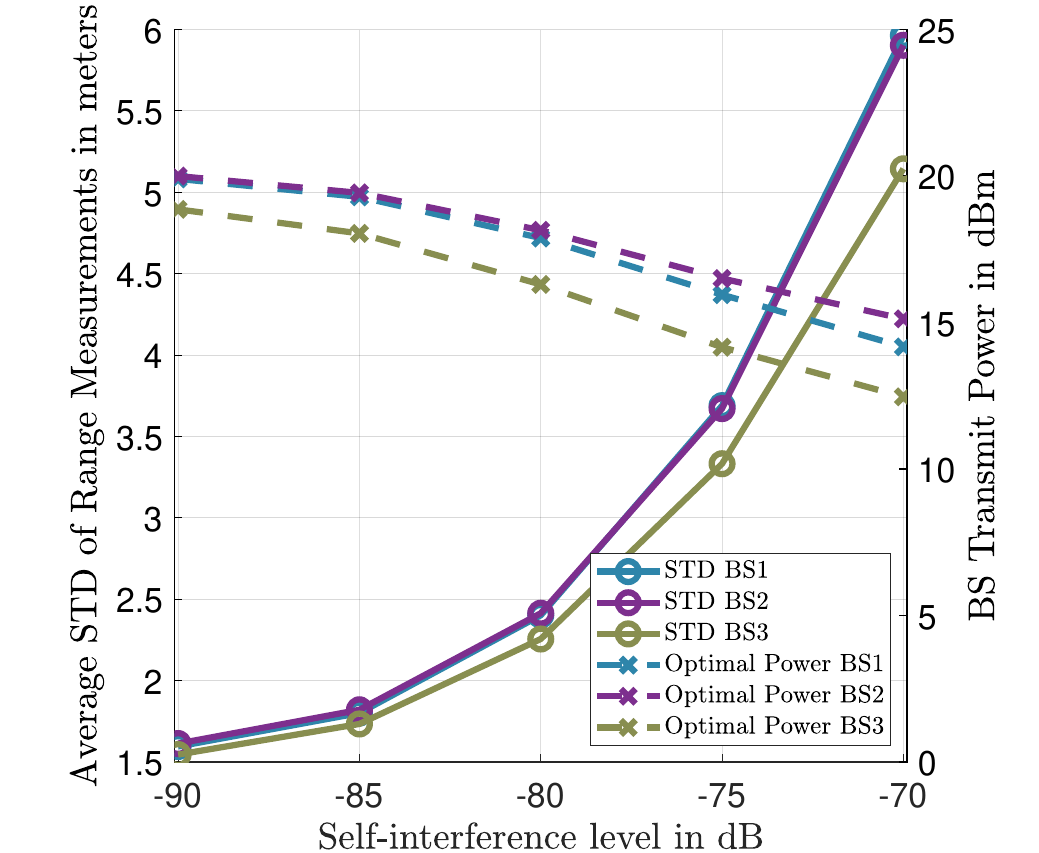}
    \caption{Average STD of range measurements of each BS vs. different SI levels with the BSs transmit power levels, with $\Gamma^{\text{comm}}_{\check m} = 0,\,\forall \check m \in \mathcal{K}$.}
    \label{fig:std_SI_Power}
     \vspace{-0.25cm}
\end{figure}

\begin{figure}[t]
    \centering
    \includegraphics[width=0.8\linewidth, trim={2cm 6.5cm 2cm 7cm},clip]{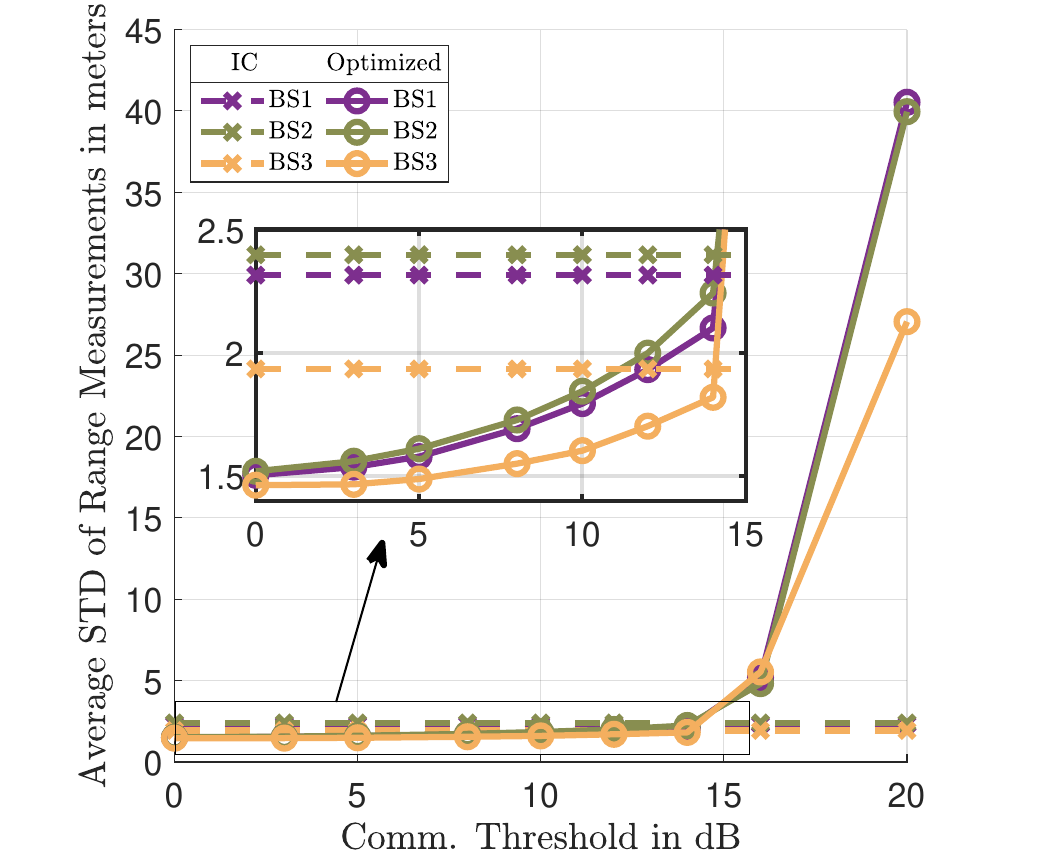}
    \caption{Average STD of range measurements of each BS vs. different communication threshold constraints, with SI $-90$ dBm.}
    \label{fig:std_CommThres}
     \vspace{-0.4cm}
\end{figure}

Fig. \ref{fig:CDF_v_STD_underSI} depicts the empirical cumulative density function (CDF) in function of the STD of range measurements ($\sigma_{R}$) in meters for different SI levels. $\sigma_{R}$ represents the average of the three BSs STD of the range measurements. This figure compares the initial condition (IC), where the power vector is set to $23$ dBm to all BSs, with the output of the proposed algorithm (named optimized in the figure). It is evident from the results that the algorithm can enhance $\sigma_{R}$ for different SI levels that are below $-80$ dB. However, beyond this level, the SI term in Eq. \eqref{eq:SINR_Sens_Ov} begins to dominate, and the interference reduction becomes less effective. Moreover, the lower the SI level is, the better the algorithm can allocate the power in the network in order to reduce the interference and increase $\sigma_{R}$.

Fig. \ref{fig:std_SI_Power} shows the average STD of range measurements for each BS ($\overline{\sigma}_{R, n},\,n \in\,[1,\,2,\,3]$) based on various SI levels and the base stations (BS) transmit power from the proposed algorithm. The power allocation scheme aims to lower the transmit power of all BSs as the SI level rises in order to contain the STD of range measurements. 
Furthermore, the level of $\overline{\sigma}_{R, n},\,n \in\,[1,\,2,\,3]$, is close to $5$ to $6$ m for SI level of $-70$ dB, indicating that the range measurements are not accurate enough. For instance, in order to achieve cooperative driving maneuver in an urban scenario, an accuracy of $1.5$ m is needed \cite{Bartoletti2021}.
Therefore, additional post-processing techniques have to be considered to estimate the range effectively, for instance, zero-Doppler cancellation, a well-developed radar technique \cite{Richard2004}. 

Fig. \ref{fig:std_CommThres} displays $\overline{\sigma}_{R, n},\,n \in\,[1,\,2,\,3]$ as a result when increasing the communication threshold constraint. This figure compares the IC with the output of the proposed algorithm. The enhanced view of the figure shows that at the lower SINR threshold range, the proposed algorithm outperforms the IC regarding sensing performance. However, when the communication service necessitates a substantial data rate under strict conditions, the algorithm ultimately compromises its sensing performance to fulfil the requisite demands. This clearly captures the tradeoff between sensing and communication performance in the ISAC networks.


\vspace{-0.2cm}
\section{Conclusions}\label{sec:Conclusions}
In this paper, we studied and evaluates the performance of joint localization and communication in monostatic sensing-based ISAC networks under the effects of network level interference as well as self-interference. We proposed a power allocation algorithm to maximize the least sensing performer BS in an ISAC network, subject to communication threshold constraints. The algorithm accounts for interference from other users, echoes from other BS, and self-interference. The proposed algorithm can enhance the sensing performance when the self-interference is effectively suppressed. 
Moreover, depending on the communication demands, the proposed algorithm can improve or decrease the sensing performance in order to meet the communication requirements. In future, we will analyze the impact of channel uncertainty and BS asynchronization on the overall localization performance in ISAC networks. 
\vspace{-0.3cm}
\small
\section*{Acknowledgment}
This work was partially funded by the Federal Ministry of Education and Research Germany within the project ”KOMSENS-6G” under grant 16KISK128 and partially funded under the Excellence Strategy of the Federal Government and the Länder under grant "RWTH-startupPD 441-23".
\vspace{-0.1cm}
\bibliographystyle{IEEEtran}
\bibliography{Power_Allocation_Multi_monostatic.bib}

\end{document}